\pdfoutput=1
\documentclass[11pt,aps,a4paper,eqsecnum,amsmath,amssymb
              ,superscriptaddress,notitlepage,nofootinbib
              ,longbibliography
              ]{revtex4-1}

\usepackage{graphicx}

\newcommand{\half}{\frac{1}{2}}

\begin{document}
\preprint{}
\title{Finite-size effects in the spectrum of the $OSp(3|2)$ superspin chain}  

\author{Holger Frahm}
\affiliation{%
Institut f\"ur Theoretische Physik, Leibniz Universit\"at Hannover,
Appelstra\ss{}e 2, 30167 Hannover, Germany}

\author{M\'arcio J. Martins}
\affiliation{%
Departamento de F\'isica, Universidade Federal de S\~ao Carlos,
C.P. 676, 13565-905 S\~ao Carlos (SP), Brazil}


\begin{abstract}
  The low energy spectrum of a spin chain with $OSp(3|2)$ supergroup symmetry
  is studied based on the Bethe ansatz solution of the related vertex model.
  This model is a lattice realization of intersecting loops in two dimensions
  with loop fugacity $z=1$ which provides a framework to study the critical
  properties of the unusual low temperature Goldstone phase of the $O(N)$
  sigma model for $N=1$ in the context of an integrable model.  Our
  finite-size analysis provides strong evidence for the existence of continua
  of scaling dimensions, the lowest of them starting at the ground state.
  Based on our data we conjecture that the so-called watermelon correlation
  functions decay logarithmically with exponents related to the quadratic
  Casimir operator of $OSp(3|2)$.  The presence of a continuous spectrum is
  not affected by a change to the boundary conditions although the density of
  states in the continua appears to be modified.

\end{abstract}

\maketitle

\section{Introduction}
\begin{figure}[h]
\includegraphics[width=0.65\textwidth]{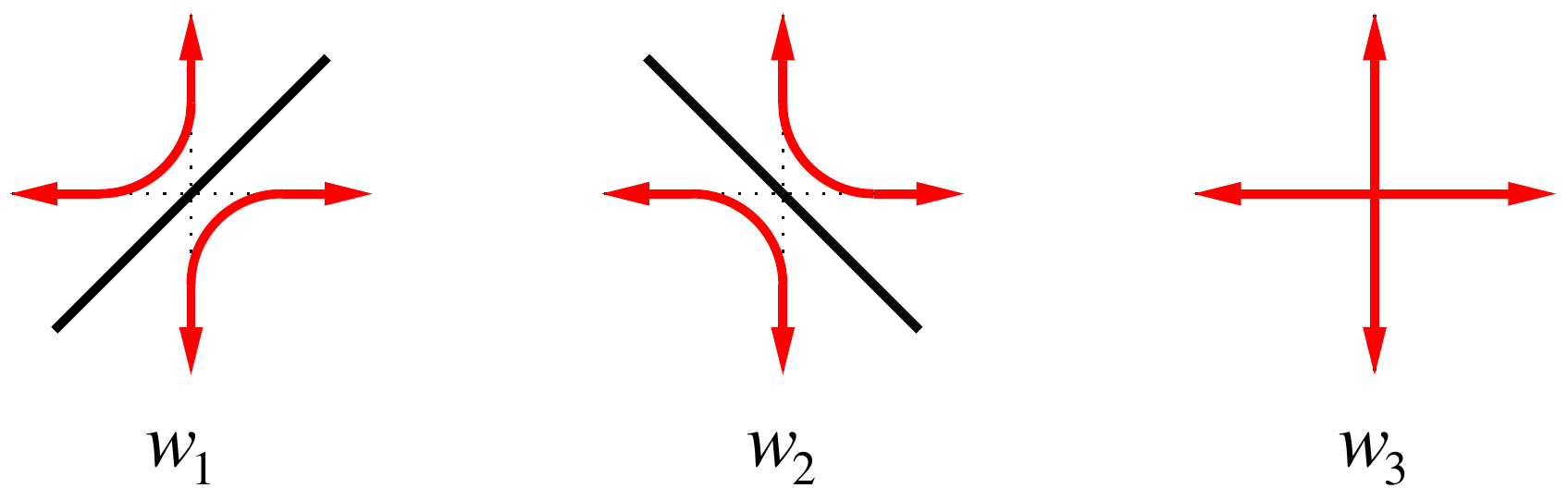}
\caption{Scattering rules for right mirrors, left mirrors and empty sites.
  The different vertices appear with probabilities $w_1$, $w_2$, $w_3$
  respectively.   \label{fig:mirrors}}
\end{figure}

This paper is concerned with the study of the finite-size properties of a
solvable two-dimensional vertex model based on the five-dimensional
representation of the $OSp(3|2)$ superalgebra. This system has a close relation
with a particular Lorentz lattice gas used to model the diffusion of particles
through randomly placed obstacles on the square lattice \cite{MaNR98}.  In
this cellular automata the particle moves along the bonds of the lattice and
is scattered according to scattering rules fixed a priori once it reaches a
given node.  Here the scatterers are constituted of mirrors tilted right and
left, i.e.\ by $\pm \frac{\pi}{4}$, with respect to the lattice
\cite{RuCo88,GuOr85}.  When the particle collides with a mirror it will turn
right or left, depending on the orientation of the latter.
In the absence of a mirror at a node the particle passes the node on a
straight path.  The corresponding scattering rules are depicted in
Figure~\ref{fig:mirrors}.
Amplitudes $w_1$, $w_2$ and $w_3$ represent the fraction of right and left
mirrors and node vacancies on the lattice, respectively.  
The kinetic properties of this lattice Lorentz gas have been investigated by
numerical simulations where an anomalous diffusive behavior was observed
\cite{ZiKC91}. In the case of partially occupied lattice by mirrors ($w_3 \ne
0$) the fractal dimension of large trajectories was argued to be $d_f=2$ with
the presence of logarithmic corrections \cite{OwPr95,CaCo97}.

\begin{figure}[h]
\includegraphics[width=0.65\textwidth]{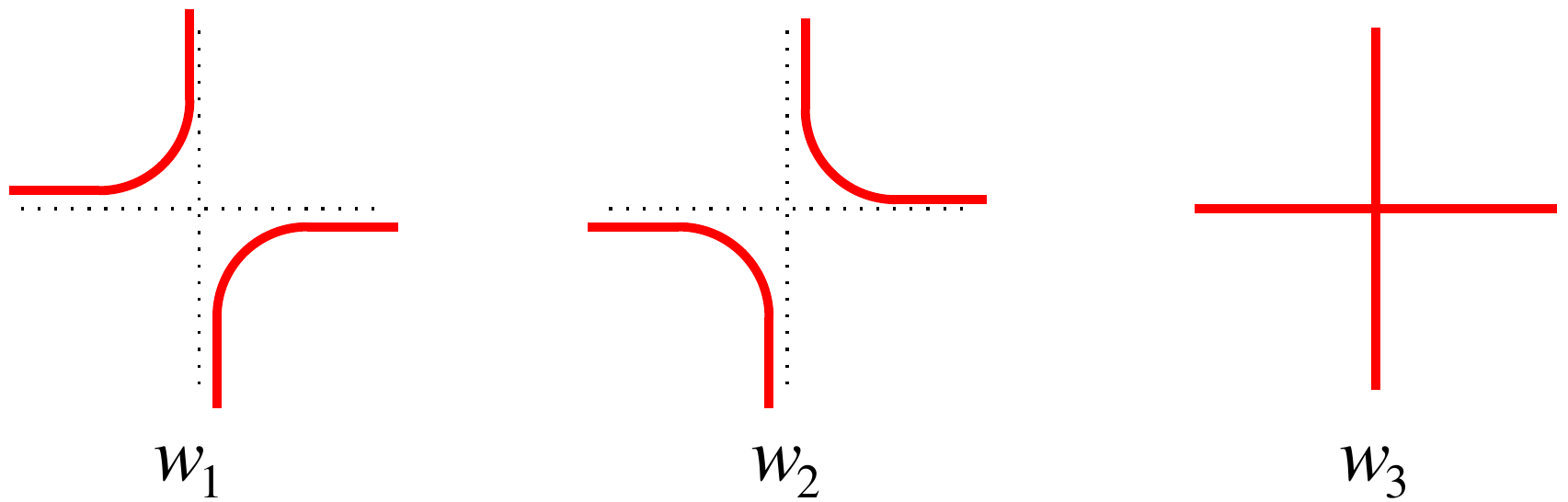}
\caption{The possible configurations of a node in the intersecting loop model
  and their associated Boltzmann weights $w_1$, $w_2$,
  $w_3$.  \label{fig:loopconf}}
\end{figure}

An alternative interpretation of the paths of particles in this lattice gas is
in terms of the degrees of freedom of an intersecting loop model with periodic
boundary conditions in both lattice directions imposed.  In
Figure~\ref{fig:loopconf} we show the three possible configurations of a node
of such loop model together with their corresponding probabilities.
Weighting every closed loop in a given state by a fugacity $z$ the object of
our study is the partition function,
\begin{equation}
\label{SUMZ}
  Z= \sum_{\mathrm{loop~configurations}} w_1^{m_1}\,w_2^{m_2}\,w_3^{m_3}\,z^{\cal{N}}\,,
\end{equation}
where $m_1$, $m_2$ and $m_3$ are the number of weights $w_1$, $w_2$ and $w_3$
and $\cal{N}$ denotes the number of loops on a such statistical configuration.

For the case of fugacity $z=1$ this intersecting loop model can be
reformulated as the $OSp(3|2)$ supersymmetric vertex model which is the subject of
this paper.  The integrability of this model provides a framework for the
study of its critical behaviour and -- exploiting the equivalences listed
above -- of the peculiar properties observed in the Lorentz lattice gas.
In fact, the finite-size analysis of the lowest excitation of the $OSp(3|2)$
superspin chain found the respective critical exponent to be very small
\cite{MaNR98}.  This was taken as an indication for the presence of a zero
conformal dimension on the spectrum implying the superdiffusive behaviour
($d_f=2$) predicted for the Lorentz gas.

In the context of the loop model, it has been argued that this behaviour
signals the existence of an unusual critical phase of intersecting loops: in
two dimensions the crossing of loops, $w_3\ne0$, is a relevant perturbation to
the low temperature dense loops phase \cite{JaRS03}.  
As a consequence the long distance behaviour of correlation functions in the
loop model with fugacity of closed loops $z=N<2$ should be that of the
Goldstone phase of the $O(N)$ sigma model.
For integer $N$ this regime can be described in a supersymmetric formulation
of the field theory in terms of $m$ bosons and $2n$ symplectic fermions,
$N=m-2n$.  The $OSp(3|2)$ vertex model is one from a class of integrable lattice
regularizations of these models.  Their central charges are $c=N-1$ as
expected in the Goldstone phase \cite{MaNR98}.

There has been a series of attempts towards the identification of the other
characteristic feature of this Goldstone phase starting from integrable
lattice models, i.e.\ a finite density of vanishing critical exponents.  The
existence of a continuous spectrum of conformal weights has been established
in several staggered superspin chains \cite{EsFS05,FrMa11,VeJS14}, including a
model based on the four-dimensional representations of $U_q[sl(2|1)]$
alternating with their duals which, in the self-dual case, is isomorphic a
deformation of the $OSp(2|2)$ chain relevant to the loop model with fugacity
$N=0$ \cite{FrMa12}.  Further details of the spectral properties of these
models have been uncovered when it was realized that hidden within the zero
charge sector of this superspin chain there exists a staggered six-vertex
model which already displays a continuous low energy spectrum.  For the latter
strong evidence has been accumulated that the effective theory describing the
low energy excitations of the model is the $SL(2,\mathbb{R})/U(1)$ sigma model
at a level related to the anisotropy \cite{IkJS08,IkJS12,CaIk13,FrSe14}.  We
note, however, that the focus of these studies has been on the anisotropic
deformations of the vertex models: while it has been established that the
isotropic $N=0$ models are on the boundary of the critical region, the
question of the critical properties at the isotropic point itself has not been
addressed.

In this paper we want to return to the case of $N=1$ as described by the $OSp(3|2)$
supersymmetric vertex model.  As argued above, the critical properties of this
model are those of the proposed Goldstone phase of the $O(1)$ sigma model and
can be studied based on its solutions by means of the algebraic Bethe ansatz.
By means of an extensive finite-size study of the model we accumulate ample
evidence for the existence of continua of critical exponents with lower edges
at scaling dimensions $X=0$, $1$, $2$, $4$, $6$,\ldots\,.  With the exception
of the the ground state of the superspin chain all of the states considered
show strong logarithmic corrections to scaling governed by the flow of the
model to weak coupling in the Goldstone phase.  This is complemented by the
observation that many excitations which have energies $\propto 1/L$ for small
system sizes but disappear from the low energy spectrum as the system size is
increased.  Both of these features of the spectrum require very large system
sizes to be studied for a reliable identification of the low energy effective
theory: here we consider lattices with up to $4096$ sites.

Finally, since recent numerical studies of the $N=1$ intersecting loop model
have emphasized the importance of boundary conditions for the long distance
behaviour of correlation functions \cite{KaNi06,NSSO13} we consider both
periodic boundary conditions to the superspin chain and twisted ones depending
on the fermion number.  Comparing the results we find that the amplitudes of
the subleading (logarithmic in the system size) finite-size corrections do
depend on the choice of boundary conditions.

\section{The integrable lattice model}
\label{sec:intmodel}
The statistical configurations of the Lorentz lattice gas or intersecting loop
model mentioned in the introduction have a one-to-one correspondence with the
generators of the braid-monoid algebra. This fact has been elaborated
previously in the work \cite{MaNR98} but for sake of completeness we have
summarized this equivalence in Appendix~\ref{app:braidmonoid}.  This algebra
can be used to built solvable models and it turns out that integrability is
assured when the weights are parameterized as
\begin{equation}
w_1=w_0,\quad w_2=\frac{w_0\lambda}{1/2-\lambda},\quad w_3=w_0\lambda
\end{equation}
where $\lambda$ is a free spectral parameter and $w_0$ is an arbitrary
normalization. This scale can be chosen to interpret the weights as 
probabilities but in our context we set it to unity.
Note that all three weights are
positive for $0\leq\lambda\leq\frac12$.

Furthermore, it has been shown that this integrable manifold can be realized
in terms of a standard local vertex model. Its bond states are constituted of
three bosonic and two fermionic degrees of freedom realized in terms of the
five-dimensional representation of the $OSp(3|2)$ superalgebra, see
Appendix~\ref{app:osp}.
The possible configurations on a vertex with their respective Boltzmann
weights are encoded in the $R$-matrix
\begin{equation}
  \label{o32rmat}
  {R}_{0j}(\lambda)= \sum_{a,b=1}^{5}\left[w_1(-1)^{p_ap_b}
    e_{ab}^{(0)} \otimes e_{ba}^{(j)}
    +w_2\sum_{c,d=1}^{5}\alpha_{ab}\alpha_{cd}^{-1}
    e_{ac}^{(0)} \otimes e_{bd}^{(j)}
    +w_3e_{ab}^{(0)} \otimes e_{ab}^{(j)}
\right]
\end{equation}
where $e_{ab}^{(k)}$ are the $5 \times 5$ Weyl matrices acting either on the
auxiliary space for $k=0$ or on the quantum space associated to the sites of a
chain of length ${L}$ for $k=1,\cdots,{L}$. The symbol $p_a$ denotes the
Grassmann parities distinguishing the bosonic ($p_a=0$) and fermionic
($p_a=1$) degrees of freedom.
The $5\times5$-matrix $\alpha_{ab}$ is the basic ingredient to built an
explicit representation for the monoid operator and its expression depends
much on the grading order basis, see for instance \cite{MaRa97a}.  Here we
will consider two specific Grassmann orderings in which the two ${U}(1)$
charges of the $OSp(3|2)$ algebra commuting with the operator ${R}_{0j}(\lambda)$
are organized in a way which is suitable to perform the Bethe ansatz analysis.
This turns out to be the ${fbbbf}$ and ${bfbfb}$ basis ordering and the
corresponding forms for the matrix $\alpha$ are
\begin{equation}
\label{alph}
\alpha^{(fbbbf)}=\left(\begin{array}{ccccc}
                0& 0 & 0 & 0 & 1 \\
                0& 0 & 0 & 1 & 0 \\
                0& 0 & 1 & 0 & 0 \\
                0& 1 & 0 & 0 & 0 \\
                -1& 0 & 0 & 0 & 0 \\
                \end{array}\right),\quad
\alpha^{(bfbfb)}=\left(\begin{array}{ccccc}
                0& 0 & 0 & 0 & 1 \\
                0& 0 & 0 & -1 & 0 \\
                0& 0 & 1 & 0 & 0 \\
                0& 1 & 0 & 0 & 0 \\
                1& 0 & 0 & 0 & 0 \\
                \end{array}\right).
\end{equation}

For the vertex model on the square lattice with $L\times L$ vertices and
periodic boundary conditions for both bosonic and fermionic configurations in
the horizontal direction we now construct the vertex model row-to-row transfer
matrix.  This operator is given as the supertrace over the auxiliary space of
an ordered product of ${L}$ matrices ${R}_{0j}(\lambda)$,
\begin{equation}
  \label{transfm}
  {T}(\lambda)=\sum_{a=1}^{5} (-1)^{p_a} 
  \left[{R}_{0{L}}(\lambda) {R}_{0{L}-1}(\lambda) 
    \cdots {R}_{01}(\lambda)\right]_{aa}\,.
\end{equation}
As a consequence of integrability the transfer matrix commutes for different
values of the spectral parameter, $[T(\lambda),T(\mu)]=0$, and therefore
generates a family of commuting operators.  As usual, a Hamiltonian with local
(nearest neighbour on the lattice) interactions is obtained by expanding
(\ref{transfm}) around the point $\lambda=0$ where the transfer matrix becomes
proportional to the translation operator.  The resulting expression for the
integrable $OSp(3|2)$ superspin Hamiltonian is
\begin{equation}
  \label{o32hamil}
  {H}=-\sum_{j=1}^{L} \left[ 
    \sum_{a,b=1}^{5}(-1)^{p_ap_b}
    e_{ab}^{(j)} \otimes e_{ba}^{(j+1)}
    +2\sum_{a,b,c,d=1}^{5}\alpha_{ab}\alpha_{cd}^{-1}
    e_{ac}^{(j)} \otimes e_{bd}^{(j+1)}\right]\,,
\end{equation}
where periodic boundary conditions for bosonic and fermionic degrees of
freedom are assumed. 

In terms of the transfer matrix the partition function of the vertex model is
given by a supertrace of the $L^\mathrm{th}$ power of $T(\lambda)$, now
taken on the $5^L$-dimensional quantum space.  For system sizes up to
$4\times4$ we find by direct computation
\begin{equation}
  \label{partfun}
  {Z}=\sum_{k=1}^{5^L} (-1)^{p_1^k+p_2^k\cdots+p_{{L}}^k} 
  {T}_{kk}(\lambda)=
  \left[1+\lambda+\frac{\lambda}{1/2-\lambda}\right]^{{L}^2}\,
\end{equation}
in agreement with the triviality of the partition sum (\ref{SUMZ}) of the loop
model with $z=1$.  Here $p_1^k,\cdots,p_{{L}}^k$ are the Grassmann parities of
the degrees of freedom composing a given $k$-state of the Hilbert space.
The partition function of any statistical model is dominated by the largest
eigenvalue of the transfer matrix.  In the present case the contribution of a
single vertex to the partition function is the sum of the three weights
$w_1+w_2+w_3$.  Therefore, as long as the Boltzmann weights are all
non-negative (i.e.\ in the regime $0 \leq \lambda \le 1/2$), we conclude that
the largest eigenvalue of the transfer matrix $\Lambda_{{max}}(\lambda)$ for
the lattice with linear dimension $L$ is
\begin{equation}
  \label{transfmax}
  \Lambda_{{max}}(\lambda)=
  \left[1+\lambda+\frac{\lambda}{1/2-\lambda}\right]^{{L}}\,.
\end{equation}
The fact that there are no subleading (in $L$) corrections is a consequence of
the grading of the states together with the properties of the $OSp(3|2)$
representations appearing in the Hilbert space, see Appendix~\ref{app:osp}.

We note that as an immediate consequence of the expression (\ref{transfmax})
for the largest eigenvalue of the transfer matrix the ground state energy of
the superspin chain (\ref{o32hamil}) is
\begin{equation}
  \label{eflat}
  E_0=-3L\,
\end{equation}
without finite-size corrections.

Eq.~(\ref{partfun}) implies that the partition function can be normalized to
$Z=1$ by rescaling of the local Boltzmann weights.  Note that this does not
necessarily mean that the low-lying excitations in the spectrum of the
transfer matrix are trivial.  In general, we can only infer that the critical
properties are governed by a conformal field theory (CFT) with central charge
$c=0$.  Since the Hamiltonian (\ref{o32hamil}) is a non-Hermitian operator the
continuum limit is not expected to be described by a \emph{unitary} $c=0$
conformal field theory.

In order to study the finite-size properties of the low-lying spectrum of this
quantum spin chain we turn to its Bethe ansatz solution.

\section{The Bethe ansatz}

The diagonalization of the transfer matrix can be carried out within the
algebraic Bethe ansatz framework.  The essential tools have already been
discussed before \cite{MaRa97a} and here we shall present only the main
results.  It turns out to be convenient to re-scale the spectral parameter by
the imaginary unit to bring the resulting equations into a canonical form for
performing numerical analysis.  Denoting by $\Lambda(\lambda)$ the eigenvalues
of the transfer matrix (\ref{transfm}) we find that they can be parametrized
in terms of two sets of rapidities $\{\lambda_j^{(a)}\}$, $a=1,2$.  The
functional form of the eigenvalues as well as the algebraic Bethe equations
satisfied by the rapidities depend on the grading chosen.

\subsection{Bethe ansatz in $fbbbf$ grading}
The Hilbert space of the superspin chain can be decomposed into the
irreducible representations (irreps) $(p;q)$ of $OSp(3|2)$ appearing in the tensor
product $(0;\frac12)^{\otimes L}$ of local spins, see Appendix~\ref{app:osp}.
In the grading $fbbbf$ a highest weight state of the $(8L-4)$-dimensional
representation $(L-1;\frac12)$ is used as pseudo vacuum (or reference state)
for the algebraic Bethe ansatz.  Thanks to the algebra inclusion ${OSp(3|2)}
\supset {SU}(2) \oplus {SU}(2)$ all states can be characterized in terms of
the two $U(1)$ charges $\ell^z$, $s^z$ from the $SU(2)$ subalgebras.  In the
reference state they take values $L$ and $0$, respectively.  Bethe states in a
$(p;q)$ multiplet are parametrized by $(L-n_1)$ complex rapidities
$\lambda_j^{(1)}$ and $(L-n_1-n_2)$ complex rapidities $\lambda_j^{(2)}$,
where
\begin{equation}
  \label{pqnos}
   n_1 = p+1\,,\quad n_2=2q-1\,
\end{equation}
for all states except the $OSp(3|2)$ singlet $(0;0)$ for which $n_1=n_2=0$
(note that only $OSp(3|2)$ irreps with integer $p$ appear in the Hilbert space of
the superspin chain).

The corresponding transfer matrix eigenvalue is given by the following
expression: 
\begin{equation}
  \begin{aligned}
    &\Lambda(\lambda)=-\left[\lambda/i-1\right]^{{L}}
    \prod_{j=1}^{{L-n_1}} \frac{\lambda-\lambda_j^{(1)}+i/2}{
      \lambda-\lambda_j^{(1)}-i/2}
    +\left[\lambda/i\right]^{{L}}
    \prod_{j=1}^{{L-n_1}} \frac{\lambda-\lambda_j^{(1)}+i/2}{
      \lambda-\lambda_j^{(1)}-i/2}
    \prod_{j=1}^{{L-n_1-n_2}} \frac{\lambda-\lambda_j^{(2)}-i}{
      \lambda-\lambda_j^{(2)}} \\
    &+
    \left[\lambda/i\right]^{{L}}\left\{
      \prod_{j=1}^{{L-n_1-n_2}} \frac{(\lambda-\lambda_j^{(2)}-i)}{
        (\lambda-\lambda_j^{(2)})}
      \frac{(\lambda-\lambda_j^{(2)}+i/2)}{
        (\lambda-\lambda_j^{(2)}-i/2)}
      +\prod_{j=1}^{{L-n_1}} \frac{\lambda-\lambda_j^{(1)}-i}{
        \lambda-\lambda_j^{(1)}}
      \prod_{j=1}^{{L-n_1-n_2}} \frac{\lambda-\lambda_j^{(2)}+i/2}{
        \lambda-\lambda_j^{(2)}-i/2} \right\}\\
    &-\left[\lambda \frac{(1/2-i\lambda)}{(-i/2+\lambda)}\right]^{{L}}
    \prod_{j=1}^{{L-n_1}}
    \frac{\lambda-\lambda_j^{(1)}-i}{\lambda-\lambda_j^{(1)}}
\,.
  \end{aligned}
\end{equation}
The eigenspectrum of the Hamiltonian can be derived from this expression giving
\begin{equation}
\label{ene1}
  E(\{\lambda_{j}^{(1)}\},\{\lambda_{j}^{(2)}\})= L-\sum_{j=1}^{{L-n_1}}
  \frac{1}{[\lambda_j^{(1)}]^2+1/4}\,. 
\end{equation}

The sets of variables $\{\lambda_{j}^{(a)}\}$ are constrained by the Bethe
ansatz equations which for the ${fbbbf}$ grading are given by
\begin{equation}
\label{bae1}
\begin{aligned}
 \left[\frac{\lambda_{j}^{(1)}+i/2}
 {\lambda_{j}^{(1)}-i/2}\right]^{{L}}&=
 \prod_{k=1}^{{L-n_1-n_2}}
  \frac{\lambda_{j}^{(1)}-\lambda_{k}^{(2)}+i/2}
 {\lambda_{j}^{(1)}-\lambda_{k}^{(2)}-i/2},\quad j=1,\cdots,{L-n_1} , 
\\
 \prod_{k=1}^{{L-n_1}}
  \frac{\lambda_{j}^{(2)}-\lambda_{k}^{(1)}+i/2}
 {\lambda_{j}^{(2)}-\lambda_{k}^{(1)}-i/2}&= 
 \prod_{k \ne j }^{{L-n_1-n_2}}
  \frac{\lambda_{j}^{(2)}-\lambda_{k}^{(2)}+i/2}
 {\lambda_{j}^{(2)}-\lambda_{k}^{(2)}-i/2}, 
 \quad j=1,\cdots,{L-n_1-n_2} \,.
\end{aligned}
\end{equation}

\subsection{Bethe ansatz in $bfbfb$ grading}
In this grading the Bethe ansatz uses a highest weight state of the
$(8L-4)$-dimensional representation $(0;L/2)$ as reference state.  States in
the sector $(p;q)$ with (\ref{pqnos}) are now parameterized by $(L-n_2-1)$
rapidities $\lambda_j^{(1)}$ and $(L-n_1-n_2)$ rapidities
$\lambda_j^{(2)}$).\footnote{%
  We use the same notation for the Bethe ansatz rapidities for both gradings.
  Therefore, whenever specific root configurations are discussed, they need to
  be seen in the context of the underlying grading.}
In terms of these parameters the corresponding transfer matrix eigenvalue is
\begin{equation}
\begin{aligned}
&\Lambda(\lambda)=\left[\lambda/i+1\right]^{{L}}
 \prod_{j=1}^{{L-n_2-1}} \frac{\lambda-\lambda_j^{(1)}-i/2}{
   \lambda-\lambda_j^{(1)}+i/2}
-\left[\lambda/i\right]^{{L}}
 \prod_{j=1}^{{L-n_2-1}} \frac{\lambda-\lambda_j^{(1)}+i/2}{
   \lambda-\lambda_j^{(1)}+i/2}
 \prod_{j=1}^{{L-n_1-n_2}} \frac{\lambda-\lambda_j^{(2)}+i}{
   \lambda-\lambda_j^{(2)}} \\
&+
\left[\lambda/i\right]^{{L}}\left\{
 \prod_{j=1}^{{L-n_1-n_2}} \frac{(\lambda-\lambda_j^{(2)}-i)}{
   (\lambda-\lambda_j^{(2)})}
 \frac{(\lambda-\lambda_j^{(2)}+i/2)}{
   (\lambda-\lambda_j^{(2)}-i/2)}
 -\prod_{j=1}^{{L-n_2-1}} \frac{\lambda-\lambda_j^{(1)}}{
   \lambda-\lambda_j^{(1)}-i}
 \prod_{j=1}^{{L-n_1-n_2}} \frac{\lambda-\lambda_j^{(2)}-3i/2}{
   \lambda-\lambda_j^{(2)}-i/2} \right\}\\
&\left[\lambda \frac{(3/2+i\lambda)}{(i/2-\lambda)}\right]^{{L}}
\prod_{j=1}^{{L-n_2-1}} \frac{\lambda-\lambda_j^{(1)}}{
  \lambda-\lambda_j^{(1)}-i}\,.
\end{aligned}
\end{equation}
The expression for energies of the Hamiltonian in this grading differs from
the previous one by an overall minus sign:
\begin{equation}
\label{ene2}
  E(\{\lambda_{j}^{(1)}\},\{\lambda_{j}^{(2)}\})
  = -L + \sum_{j=1}^{{L-n_2-1}} \frac{1}{[\lambda_j^{(1)}]^2+1/4}
\end{equation}

The Bethe equations for the rapidities $\lambda_{j}^{(a)}$, $a=1,2$ in the
grading $bfbfb$ are 
\begin{equation}
\label{bae2}
\begin{aligned}
 \left[\frac{\lambda_{j}^{(1)}+i/2}
 {\lambda_{j}^{(1)}-i/2}\right]^{{L}}&=
 \prod_{k=1}^{{L-n_1-n_2}}
  \frac{\lambda_{j}^{(1)}-\lambda_{k}^{(2)}+i/2}
 {\lambda_{j}^{(1)}-\lambda_{k}^{(2)}-i/2},\quad j=1,\cdots,{L-n_2-1}, 
\\
 \prod_{k=1}^{{L-n_2-1}}
  \frac{\lambda_{j}^{(2)}-\lambda_{k}^{(1)}+i/2}
 {\lambda_{j}^{(2)}-\lambda_{k}^{(1)}-i/2}&= 
 \prod_{k \ne j }^{{L-n_1-n_2}}
  \frac{(\lambda_{j}^{(2)}-\lambda_{k}^{(2)}-i/2)}
 {(\lambda_{j}^{(2)}-\lambda_{k}^{(2)}+i/2)} 
  \frac{(\lambda_{j}^{(2)}-\lambda_{k}^{(2)}+i)}
 {(\lambda_{j}^{(2)}-\lambda_{k}^{(2)}-i)}, 
 \quad j=1,\cdots,{L-n_1-n_2} 
\end{aligned}
\end{equation}
Note that in this solution the inclusion ${OSp(3|2)} \supset {OSp(1|2)}$
becomes manifest: the second Bethe equations are exactly those of the
${OSp(1|2)}$ invariant vertex model in the presence of inhomogeneities
\cite{MaRa97a}.

\subsection{Example: $L=2$}
For $L=2$ the spectrum of the superspin chain decomposes into the $OSp(3|2)$
singlet $(0;0)$ and two twelve-dimensional multiplets $(0;1)$ and $(1;\half)$,
see Eq.~(\ref{o32-decomp}).  For these we can identify the corresponding Bethe
root configurations in both gradings:
   
For the $fbbbf$ grading the Bethe equations are (\ref{bae1}), the
corresponding energy is (\ref{ene1}).  The three energies are
\begin{itemize}
\item $(1;\frac12)$ (the $fbbbf$ reference state): \newline
  $L-n_1=0$ Bethe roots $\lambda^{(1)}$, $L-n_1-n_2=0$ Bethe roots
  $\lambda^{(2)}$: the energy of this state is  $E=+2$.
\item $(0;1)$:\newline here we have $n_1=1$, $n_2=1$ according to
  (\ref{pqnos}).  The only one finite solution to the Bethe equations
  (\ref{bae1}) is $\lambda^{(1)}=0$ giving energy $E=-2$.
\item $(0;0)$: \newline for the singlet ground state the quantum numbers are
  $(n_1,n_2)=(0,0)$, therefore there are two roots $\lambda_j^{(a)}$ on each
  level, $a=1,2$.  The unique solution to the Bethe equations is
  $\lambda_{1,2}^{(1)}=0$ (degenerate roots) and $\lambda_{1,2}^{(2)}=\pm
  i\frac{\sqrt{3}}{6}$.  The resulting energy is $E=-6$, as expected from the
  general considerations in Section~\ref{sec:intmodel}.
\end{itemize}

For the grading $bfbfb$ the Bethe equations (\ref{bae2}) have to be solved.
\begin{itemize}
\item $(0;1)$ (the $bfbfb$ reference state) :\newline the number of Bethe
  roots is $L-n_2-1=0$ and $L-n_1-n_2=0$, respectively.  With this we find
  $E=-2$, as in the other grading.
\item $(1;\frac12)$:\newline With $n_1=2$, $n_1=0$ there is only one finite
  solutions $\lambda^{(1)}=0$ giving $E=+2$.
\item $(0;0)$: \newline the singlet ground state is again parametrized by two
  roots on each level, which have to satisfy (\ref{bae2}) in this grading.  It
  is straightforward to find the solution to be $\lambda_{1,2}^{(1)}=\pm
  i\frac{\sqrt{3}}{2}$, $\lambda_{1,2}^{(2)}=0$ (degenerate roots) giving the
  ground state energy $E=-6$.
\end{itemize}

Note that the degeneration of Bethe roots in the 'flat' ground state is a
feature which has also been observed in other (super-)spin chains
\cite{AlMa90,EsFS05,FrMa11}.

\section{Ground state and lowest excitations}
As discussed above the ground state energy of the superspin chain is exactly
given by (\ref{eflat}).  For $L=2$ the corresponding root configurations in
the Bethe ans\"atze (\ref{bae1}) and (\ref{bae2}) have been obtained in the
previous section where they were found to be singular in the sense that Bethe
roots may degenerate.  For odd chain lengths $L$ the situation turns out to be
easier: here we find that the root configurations describing the $(0;\frac12)$
ground state are non-degenerate.  In the grading $fbbbf$ it is given by
collections of $(L-1)/2$ pairs of complex conjugate rapidities
\begin{equation}
  \label{bae1-pairs}
  \lambda_j^{(a)} \simeq \xi_j^{(a)} \pm \frac{i}{4}\,,\quad
  \xi_j^{(a)}\in\mathbb{R}\,,
\end{equation}
on each level $a=1,2$.  In the thermodynamic limit, $L\to\infty$, the
deviations from these 'strings' become small.  This allows to compute the
ground energy density and the Fermi velocity $v_F$ of the gapless low lying
excitations within the root density approach \cite{YaYa69} giving
$\lim_{L\to\infty}E_0(L)/L=-3$ and $v_F=2\pi$, see also
Ref.~\onlinecite{MaNR98}.

As a consequence of conformal invariance the leading terms in the finite-size
scaling of energy levels are predicted to be \cite{BlCN86,Affl86}
\begin{equation}
\label{cft:fs}
\begin{aligned}
  E_0(L) -L\epsilon_\infty &= -\frac{\pi v_F}{6L}\, c+o\left(\frac1{L}\right)\,,\\
  E_n(L) - E_0(L) &= \frac{2\pi v_F}{L}\, X_n +o\left(\frac1{L}\right)\,.
\end{aligned}
\end{equation}
Here $c$ is the central charge of the effective low energy theory.  As a
consequence of (\ref{eflat}) we have $c=0$ without corrections to scaling for
the superspin chain.  From the scaling dimensions $X_n = h+\bar{h}$ appearing
in the low energy spectrum together with the conformal spin, $s=h-\bar{h}$,
which can be read off the momentum of the corresponding state, we can obtain
the conformal weights $(h,\bar{h})$ of the operators in this CFT.
We note that the CFT with $c=0$ is not unitary which may lead to subleading
finite-size corrections vanishing as inverse powers of $\log L$ in
(\ref{cft:fs}).  The identification of these terms is one of the goals of this
work.

\begin{figure}[h]
(a) \includegraphics[width=0.65\textwidth]{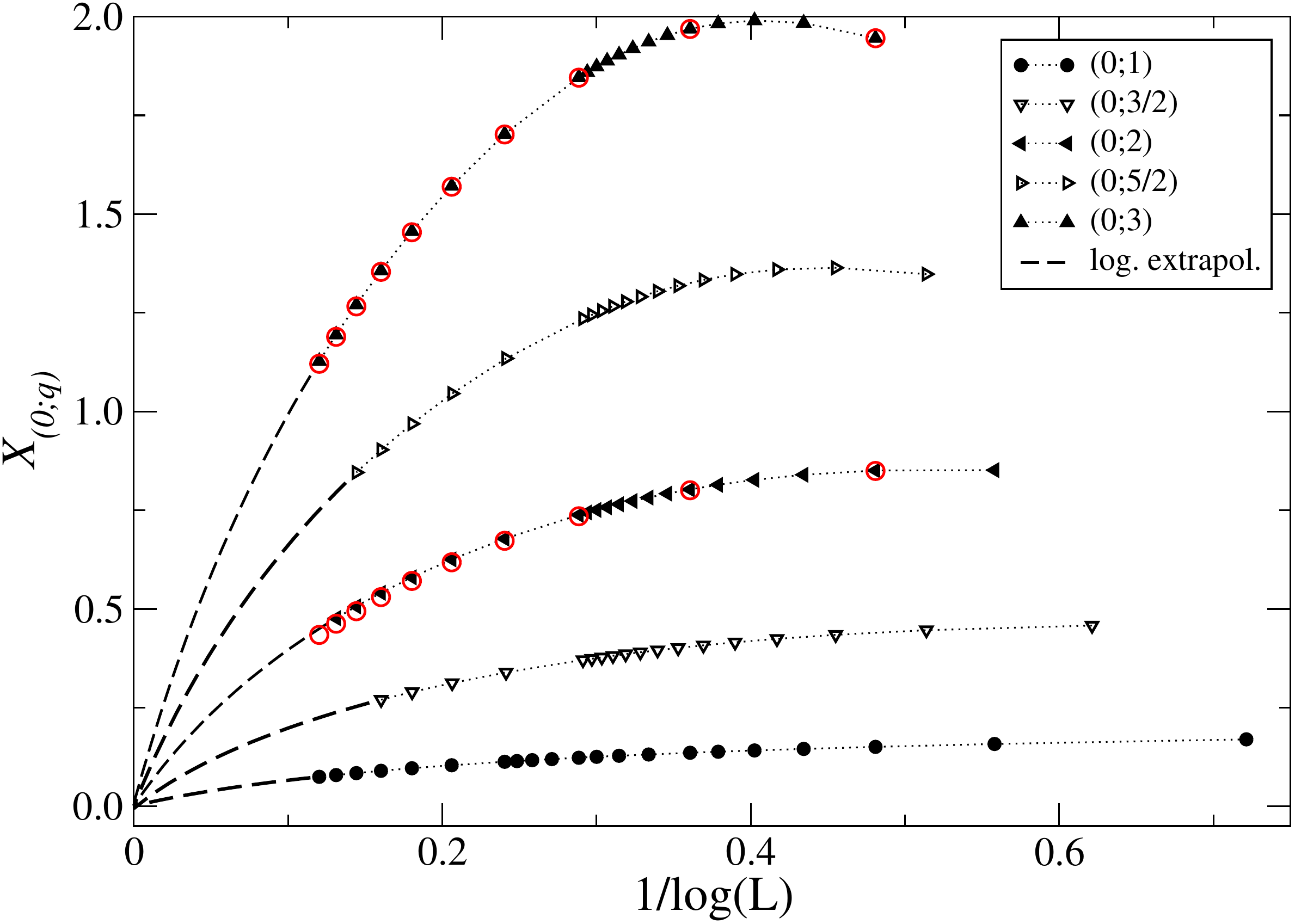}

(b)\includegraphics[width=0.65\textwidth]{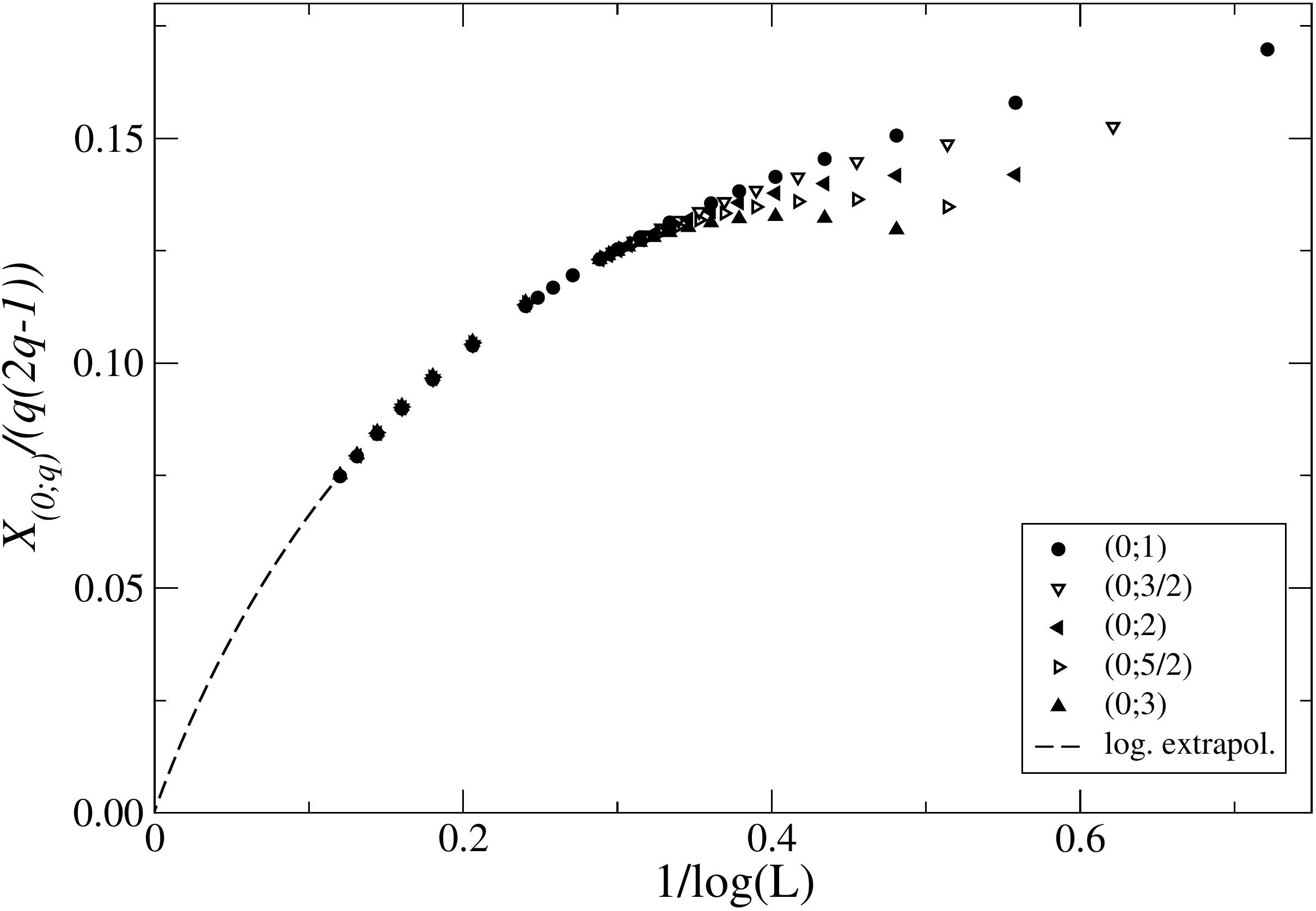}

\caption{Scaling dimensions $X_{(0;q)}$ extracted from the finite-size
  behaviour of the energies of the lowest states in the sector $(0;q)$.  These
  states have conformal spin $0$.  Data
  obtained for the superspin chain with even (odd) number $L$ of sites are
  represented by filled (open) symbols.
  The dashed lines show  the results of an extrapolation of the finite-size
  data (for system sizes up to $L=4096$) assuming a rational dependence on
  $1/\log L$, red circles indicate the results obtained within the string
  approximation (\ref{betheSTR})--(\ref{eneSTR}).
  In (b) the data for the scaling dimensions are shown with the $X_{(0;q)}$
  rescaled with the conjectured amplitude (\ref{fuseau-conj}).
  \label{fig:fuseau}}
\end{figure}

From the exact diagonalization of the superspin chain Hamiltonian for small
system sizes we find that among the lowest excitations of the superspin chain
there are the minimum energy states in the symmetry sectors corresponding to
the $(0;q)$ irreps of $OSp(3|2)$ with $q\ge1$.
The highest weight state in these multiplets reached in the first Bethe ansatz
(grading $fbbbf$) is described by $(L-n_1,L-n_1-n_2)=(L-1,L-2q)$ roots of
(\ref{bae1}).  It turns out, however, that for this class of states it is
\emph{much} easier to work within the $bfbfb$ grading, Eqs.~(\ref{bae2}).
Here the highest weight state has $(L-n_2-1,L-n1-n2)=(L-2q,L-2q)$ roots
$\lambda_j^{(a)}$ arranged into collections of complex conjugate pairs, namely
\begin{equation}
  \label{bae2-pairs}
  \lambda_j^{(1)} \simeq \mu_j^{(1)} \pm \frac{3i}{4}\,,\quad
  \lambda_j^{(2)} \simeq \mu_j^{(2)} \pm \frac{i}{4}\,,\quad
  \mu_j^{(a)}\in\mathbb{R}\,.
\end{equation}
Configurations of this type with $2q$ even (odd) can appear for $L$ even
(odd).  The numerical estimates for the scaling dimensions obtained from
solutions to the Bethe equations are slowly decreasing with the system size
once $L$ is sufficiently large.  For the lowest level ($q=1$) this has been
observed already in in Ref.~\onlinecite{MaNR98} where this was taken as
evidence for the presence of a zero conformal weight in the spectrum.
Later, this feature has been argued to be one of the characteristics of the
low temperature Goldstone phase of the $O(N)$ sigma-model with $N<2$ realized
in the presence of loop crossings \cite{JaRS03}: the latter act as a
perturbation which breaks the symmetry from $O(N)$, or $OSp(m|2n)$ with
$N=m-2n$, to $OSp(m|2n)/OSp(m-1|2n)$.  In the present case with $N=1$, and
using the system size $L$ as a long distance cutoff, the single coupling
constant of the resulting sigma model on this supersphere is found to be
\begin{equation}
  \label{gsigma}
  g_\sigma \sim \frac{1}{\log(L/L_0)}\,
\end{equation}
within a perturbative RG approach \cite{Polyakov75,ReSa01} (note that $\log
L_0$ has to be negative for  $g_\sigma\geq0$ as expected on physical
grounds).  
With this as an input we extrapolate the finite-size data for the scaling
dimensions assuming a rational dependence on $1/\log L$.  Our results for the
lowest $(0;q)$ states are displayed in Figure~\ref{fig:fuseau}.
From these data we conclude that all dimensions $X_{(0;q)}$ with finite
$q\ge1$ vanish in the thermodynamic limit, showing subleading scaling
corrections proportional to $1/\log L$.  For $q>2$ the fact that
$X_{(0;q)}\to0$ can be obscured by the latter for quite large $L$ rendering a
finite-size analysis based on small system sizes impossible.

To determine the amplitude of these subleading terms we note that this class
of $(p=0;q)$-states can be extended to include the ground states of the
superspin chain, i.e.\ the singlet $(0;0)$ for $L$ even and the quintet
$(0,\frac12)$ for $L$ odd.  Since there are no finite-size corrections to the
energies of the ground states (with $q=0$ and $q=\frac12$ for $L$ even and
odd, respectively) we conjecture that the amplitudes of the logarithms are
related to the quadratic Casimir (\ref{o32-casimir}) of $OSp(3|2)$ as
\begin{equation}
\label{fuseau-conj}
  X_{(0;q)} = 0 + \frac{q(2q-1)}{\log(L/L_0)}\,,\qquad
   q=0,\frac12,1,\frac32,2,\ldots
\end{equation}
Comparing this conjecture with our numerical data we find very good agreement,
see Figure~\ref{fig:fuseau}(b).  The lattice sizes $L$ considered here do not
allow for a realiable estimate of the non-universal scale $\log L_0$ though.
In the context of the loop model the amplitudes (\ref{fuseau-conj}) determine
the long distance asymptotics of the 'watermelon' correlation functions
$G_k(r)$ measuring the probability of $k$ loop segments connecting two points
at distance $r$: these correlators -- two-point functions of the so-called
$k$-leg operators -- vanish with a power of $1/\log r$, i.e.\
\begin{equation}
  G_k(r) \sim 1/\left(\log r\right)^{\alpha_k}\,.
\end{equation}
The amplitudes conjectured for the periodic superspin chain imply that the
exponents $\alpha_k$ are to be taken from the set $\{2q(2q-1),
q=1,3/2,2,\ldots\}$, i.e.\ the eigenvalues of the quadratic Casimir
(\ref{o32-casimir}) of $OSp(3|2)$.  Identifying the $(0;q)$ primary with the
$k=2q$-leg operator this agrees with RG calculations and numerical results for
$G_2(r)$ and $G_4(r)$ \cite{NSSO13}.  As will be seen below, however, the
asymptotic behaviour of the watermelon correlators is likely to depend on the
boundary conditions.


We further note that the reference state in the second Bethe ansatz is a
highest weight state in the unique $(0;L/2)$ multiplet of the $OSp(3|2)$ superspin
chain of length $L$.  This indicates that the spectrum of the $(0;q)$ states
considered here extends from the ground state energy $E_0=-3L$ to the energy
$E(0;L/2)=-L$.  This suggests that the critical dimensions (\ref{fuseau-conj})
actually form a continuum starting at $X_{(0;0)}=0$.
Additional support for this suggestion comes from the observation that for $q$
sufficiently large the imaginary parts of the Bethe roots in the $bfbfb$
grading are exponentially (in $L$) close to the hypothesis (\ref{bae2-pairs}).
In addition the real parts of the $(L-2q)/2$ pairs on each of the two levels
take values very close to each other, i.e. $\mu_j^{(1)} \approx \mu_j^{(2)}$.
In this situation the second set of the Bethe equations (\ref{bae2}) is
automatically satisfied while the first level ones become, after taking their
logarithm,
\begin{equation}
  \label{betheSTR}
  {L}\left[\psi_{5/4}(\mu_j)-\psi_{1/4}(\mu_j)\right]
  =-2\pi Q_j +\sum_{k \ne j}^{(L-2q)/2}
  \left[\psi_{3/2}(\mu_j-\mu_k) +
    \psi_{1}(\mu_j-\mu_k)-\psi_{1/2}(\mu_j-\mu_k) \right]\,.
\end{equation}
Here $\psi_a(x)=2\arctan(x/a)$ and the numbers $Q_j$ define the many possible
branches of the logarithm being given by the expression,
\begin{equation}
Q_j=-\frac14(L-2q) +j-\frac12,\quad j=1,\cdots,{\frac12(L-2q)}\,.
\end{equation}
Within this approach the eigenenergies corresponding to this state can be
obtained using the following expression
\begin{equation}
  \label{eneSTR}
  E(\{\mu_j\})= -L + \frac{1}{2}\sum_{j=1}^{(L-2q)/2} 
  \left[\frac{5}{(\mu_j)^2+(5/4)^2}
    -\frac{1}{(\mu_j)^2+(1/4)^2} \right]\,.
\end{equation}

Comparing the numerical solution of these 'string' equations with the those of
(\ref{bae2}) we conclude that the finite-size energies in the sectors $(0;q)$
with $q>2$ are reproduced by Eqs.~(\ref{betheSTR})--(\ref{eneSTR}), see
Figure~\ref{fig:fuseau}(a).  In this formulation the root density approach can
be applied to compute the finite-size energies.  In this approach we find
again that for $L\to\infty$ the conformal dimensions are zero, independent of
$q$.  This observation provides an additional analytical support to the
existence of a continuum of zero conformal weights starting at zero.

\section{Other excitations}
\begin{figure}[h]
\includegraphics[width=0.65\textwidth]{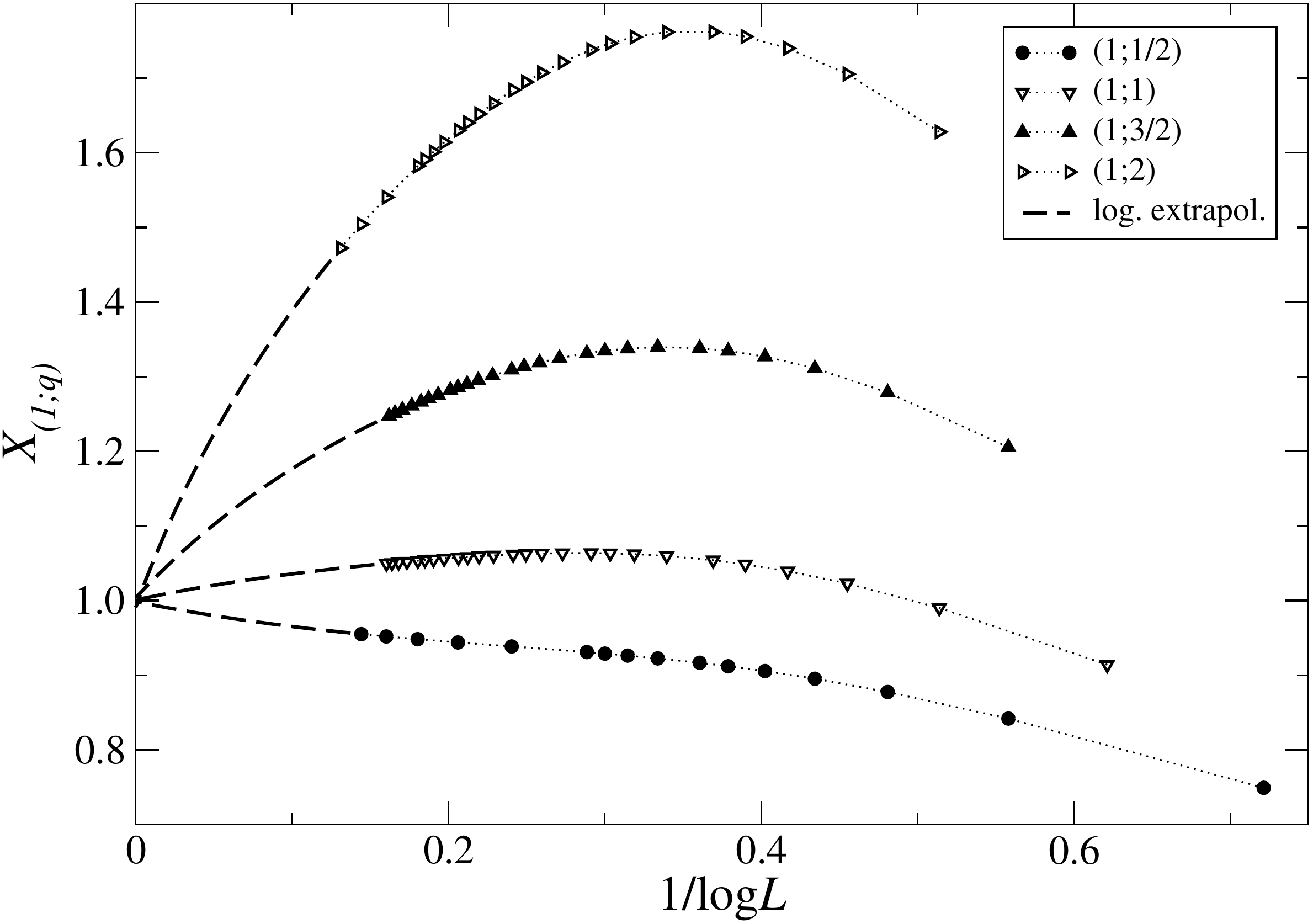}

\caption{Same as Fig.~\ref{fig:fuseau}(a) but now for  the lowest states in
  the sector $(1;q)$.  These states have conformal spin $1$.
  \label{fig:x1-q}}
\end{figure}

As for the sectors $(0;q)$ in the previous section we have identified the
Bethe configurations corresponding to the lowest energy states in the sectors
$(p;q)$ with $p\ne0$.  In Figure~\ref{fig:x1-q} we present the corresponding
scaling dimensions for $p=1$.
Again, the energies show strong logarithmic corrections to scaling which are
dealt with in the extrapolation by assuming a rational dependence of the
finite-size data on $1/\log L$.  For all states which we have considered the
scaling dimension extrapolates to $X_{(1;q)}=1$.  All of these states are
found to carry momentum $\pm2\pi/L$, from which we conclude that the
corresponding fields have spin $1$.

\begin{figure}[h]
\includegraphics[width=0.65\textwidth]{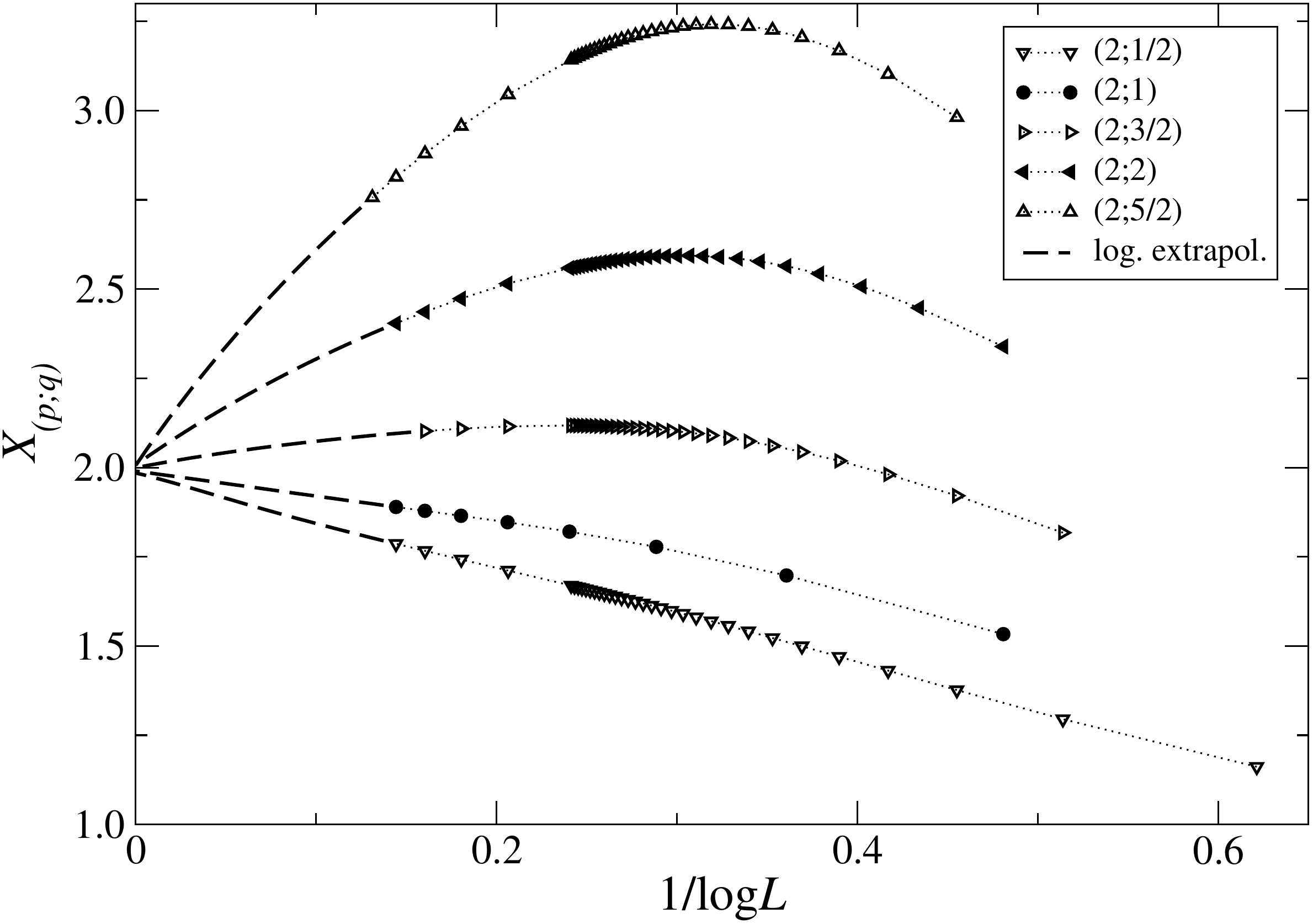}

\caption{Same as Fig.~\ref{fig:fuseau}(a) but now for  the lowest states in
  the sector $(2;q)$.  These states have conformal spin $0$.
  \label{fig:x2-q}}
\end{figure}

\begin{figure}[h]
\includegraphics[width=0.65\textwidth]{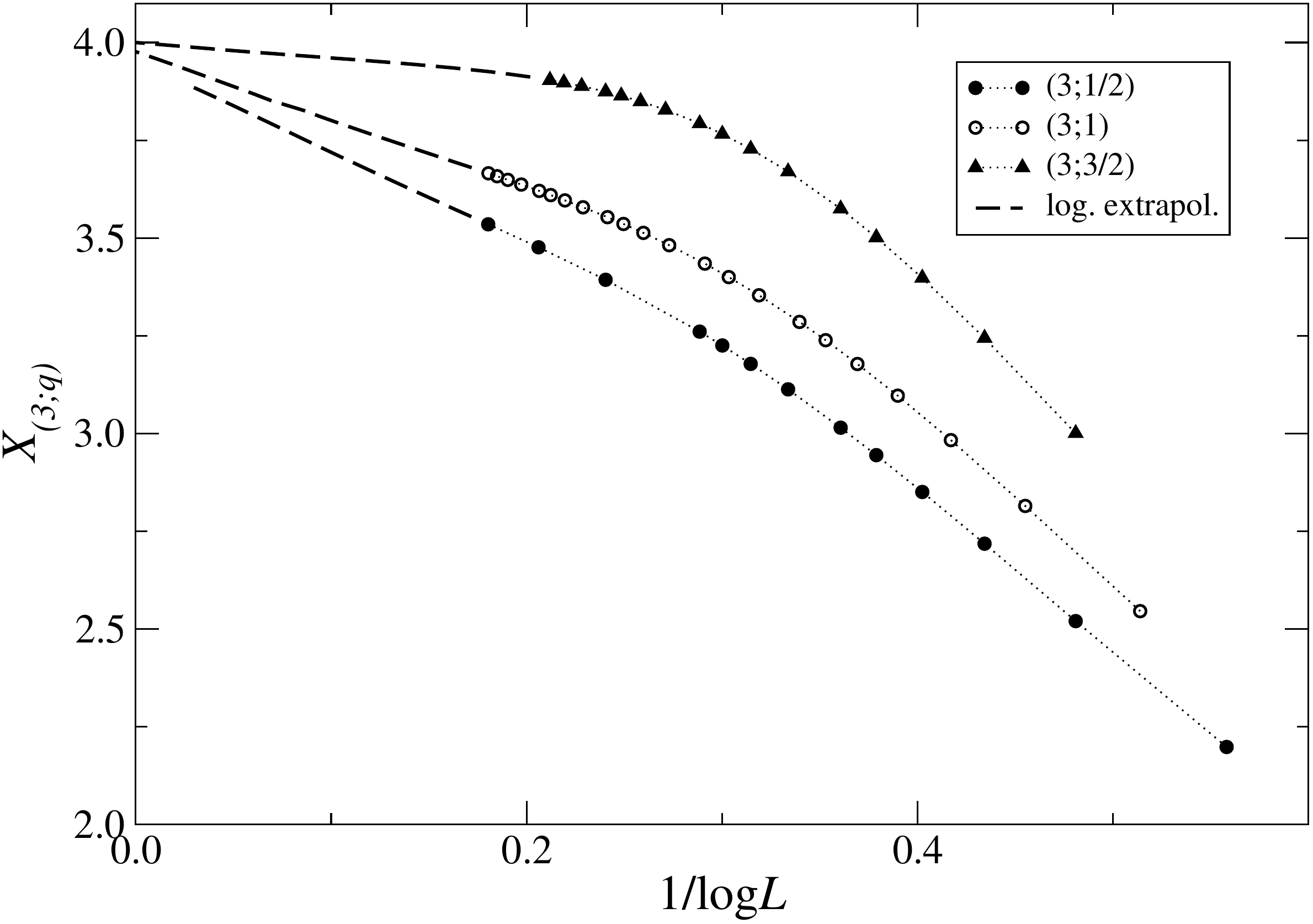}

\caption{Same as Fig.~\ref{fig:fuseau}(a) but now for  the lowest states in
  the sector $(3;q)$.  These states have conformal spin $2$.
  \label{fig:x3-q}}
\end{figure}

\begin{figure}[h]
\includegraphics[width=0.65\textwidth]{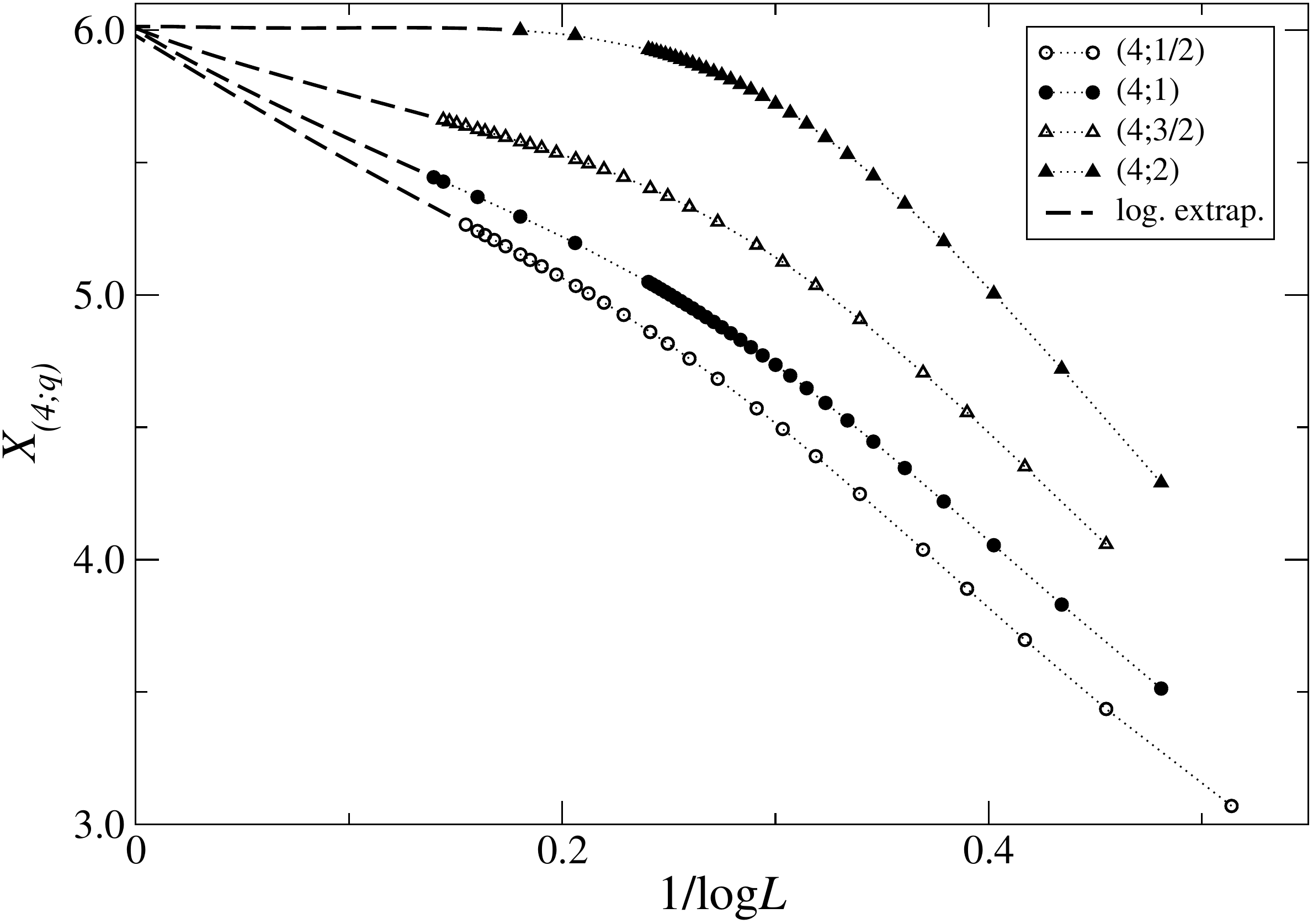}

\caption{Same as Fig.~\ref{fig:fuseau}(a) but now for  the lowest states in
  the sector $(4;q)$.  These states have conformal spin $0$.
  \label{fig:x4-q}}
\end{figure}

The lowest energy levels in some sectors $(p;q)$ with $p>2$ are studied in an
analogeous way, see Figures~\ref{fig:x2-q}, \ref{fig:x3-q}, and
\ref{fig:x4-q}.
The extrapolation gives scaling dimensions $X_{(2;q)}=2$, $X_{(3;q)}=4$, and
$X_{(4;q)}=6$ as lower edges of continua of states with different $q$.  The low
lying states in sectors with $p=2,4$ have zero momentum while the states in
the sectors $(3;q)$ have spin $2$.  Note that the presence of these levels in
the spectrum of the superspin chain of length $L$, requires the selection rule
$L-p-2q\in2\mathbb{N}$ to be satisfied.

In Table~\ref{tab:cweights} we have collected the discrete parts of the
conformal weights of primary operators in the CFT as identified from our
numerical studies.
\begin{table}[h]
  \caption{Discrete part of the scaling dimensions $X$, conformal spin $s$ and
    corresponding conformal weights $(h,\bar{h})$ obtained from the lowest
    states observed in the symmetry sectors $(p;q)$ of the superspin
    chain.\label{tab:cweights}} 
\begin{ruledtabular}
\begin{tabular}{ccccl}
  sector $(p;q)$ & $X=h+\bar{h}$ & $s=h-\bar{h}$ & $(h,\bar{h})$ \\ \hline
  $(0;q)$ & $0$ & $0$    & $(0,0)$\\
  $(1;q)$ & $1$ & $\pm1$ & $(1,0)$, $(0,1)$\\
  $(2;q)$ & $2$ & $0$    & $(1,1)$\\
  $(3;q)$ & $4$ & $\pm2$ & $(3,1)$, $(1,3)$\\
  $(4;q)$ & $6$ & $0$    & $(3,3)$\\
\end{tabular}
\end{ruledtabular}
\end{table}
Based on these data we conjecture that the spectrum of conformal weights is
given by $h_{k}=\frac12 k(k+1)$, $k=0,1,2,\ldots$ and the lowest levels in
the $(p;q)$ sector of the spectrum of the superspin chain correspond to
operators with conformal weights
\begin{equation}
  \left(h,\bar{h}\right) =
    \begin{cases}
      \left(h_{p/2},h_{p/2}\right)& \mathrm{for~} p\mathrm{~even}\,,\\
      \left(h_{(p\pm1)/2},h_{(p\mp1)/2}\right)& \mathrm{for~} p\mathrm{~odd}\,.
    \end{cases}
\end{equation}

\begin{figure}[h]
\includegraphics[width=0.65\textwidth]{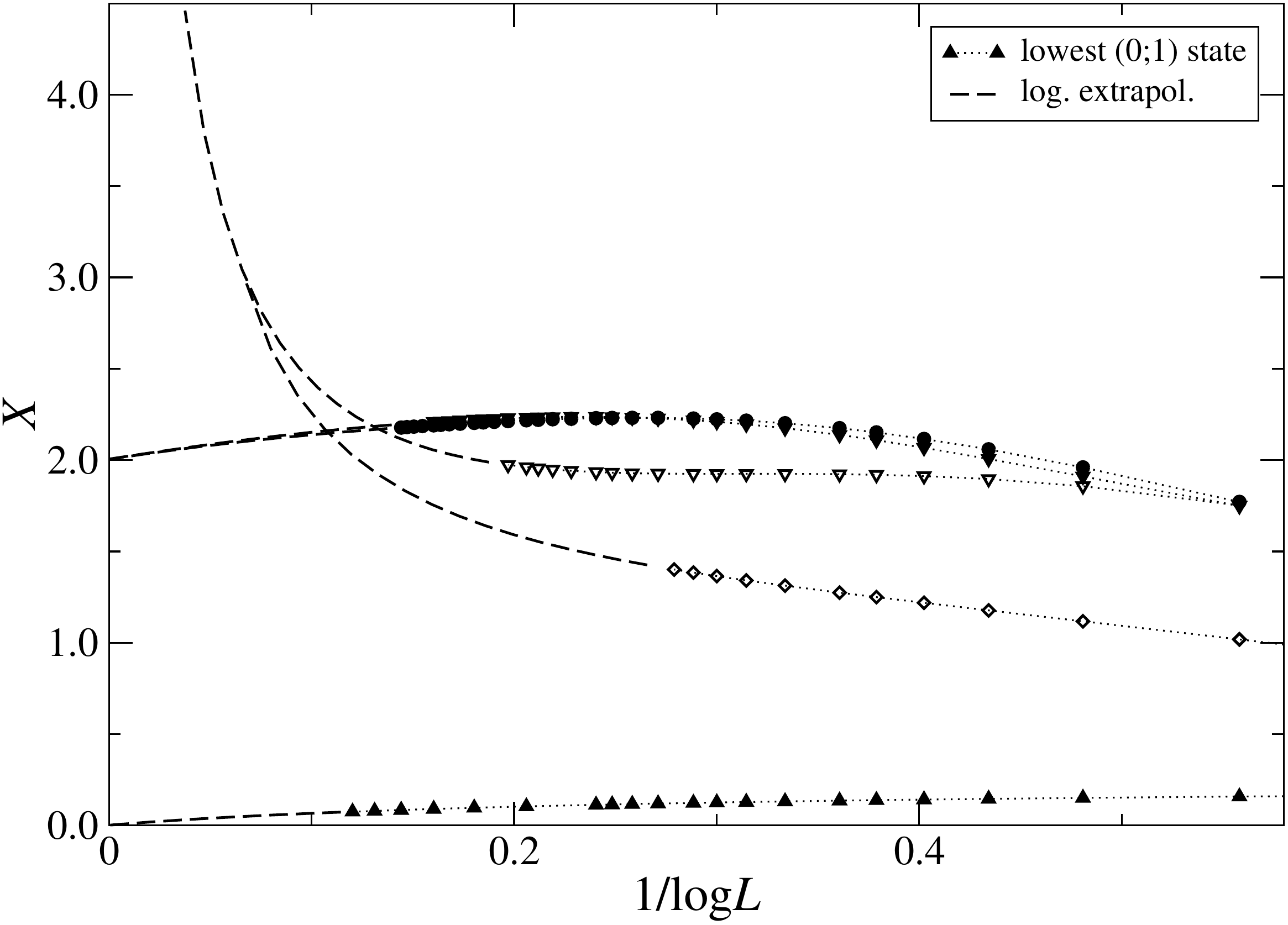}
\caption{Scaling dimensions $X$ extracted from the finite-size behaviour of
  the energies of some low lying states in the $(0;1)$ sector together with
  their extrapolation assuming a rational dependence on $1/\log
  L$.  The levels with $X\sim 1$ ($1.75$) for small systems (largest $1/\log
  L$) have conformal spin $1$ and $0$, respectively. \label{fig:pq_0-10}}
\end{figure}

Having identified the primary operators responsible for the lowest energy
states we proceed to studying the finite-size behaviour of the other states
appearing in the sector $(0;1)$:
from the numerical diagonalization of the Hamiltonian (\ref{o32hamil}) for the
superspin chain with $L=6$ sites we observe, apart from the $(0;1)$ primary
with vanishing scaling dimension (\ref{fuseau-conj}), two low lying
excitations with spin $\pm1$ and three low lying excitations with zero spin.
We have identified the configurations of Bethe roots corresponding to all of
these except one of the spin $1$ states.
Solving the Bethe equations for these configurations for larger system sizes
and computing the corresponding scaling dimensions we find that they
extrapolate to $X=2$ for two of the spin $0$ states, indicating a descendent
field of the $(0;1)$ primary operator.  The scaling dimensions of the third
spin $0$ state and the spin $1$ state, however, do not converge to a finite
value under extrapolation assuming a rational dependence on $1/\log L$ but
rather disappear from the low energy spectrum, see Figure~\ref{fig:pq_0-10}.

This behaviour is not limited to excitations in the $(0;1)$ sector but we have
found such levels in many other sectors, too.  Let us note that the removal of
low energy states present in small systems in the scaling limit has also been
observed in one phase of the $U_q[sl(2|1)]$ staggered superspin chain
\cite{FrMa12}.  It is a direct consequence of the vanishing of a coupling
constant, such as in (\ref{gsigma}) for the $OSp(3|2)$ model, and therefore
expected to be a generic property of realizations for field theories with a
continuous spectrum of critical exponents as lattice models with a compact
quantum space.

From conformal field theory one expects a spin $1$ descendent of the $(0;1)$
primary field with scaling dimension $X=1$.  This might be the other low
energy spin $1$ level observed in the $L=6$ spectrum.  Since we have not able
to identify a solution to the Bethe equations (\ref{bae1}) or (\ref{bae2})
reproducing the numerical value for the energy, we cannot confirm this
expectation.

\section{Boundary Conditions}

Here we would like to point out that peculiar finite-size behavior reported so
far is also present if we had modified the boundary conditions such that the
five possible states on the bonds of the vertex model are considered to be
bosonic degrees of freedom.  In this situation the vertex model transfer
matrix is given, instead of (\ref{transfm}), as the standard trace over the
auxiliary space of the following product of operators,
\begin{equation}
\label{tm-bos}
  \bar{{T}}(\lambda)=\sum_{a=1}^{5} 
  \left[\bar{{I}}_{0{L}}{R}_{0{L}}(\lambda) 
    \bar{{I}}_{0{L}-1}{R}_{0{L}-1}(\lambda) 
    \cdots \bar{{I}}_{01}{R}_{01}(\lambda)\right]_{aa}
\end{equation}
where the operator $\bar{{I}}_{0,j}=\sum_{a,b=1}^{5} (-1)^{p_ap_b}
e_{aa}^{(0)} \times e_{bb}^{(j)}$ plays the role of a graded identity matrix.
We recall here that this type of transfer matrix has been considered before in
the case a vertex model based on the ${OSp}(1|2)$ superalgebra \cite{Mart95}.

From the point of view of the spin chain this modification corresponds to
imposing anti-periodic boundary for the fermionic degrees of freedom which
splits the $(p;q)$ multiplets of the superspin chain into two subsets
$(p;q)_e$ and $(p;q)_o$ containing states with even and odd number of
fermions, respectively.
In the Bethe equations this twist is reflected by additional phase factors
depending on the sector label of the conserved charges.  In the ${fbbbf}$
grading only the first set of the Bethe equations (\ref{bae1}) is affected by
this change of boundary conditions giving
\begin{equation}
  \label{bae1-bos}
  \begin{aligned}
    \left[\frac{\lambda_{j}^{(1)}+i/2}
      {\lambda_{j}^{(1)}-i/2}\right]^{{L}}&=(-1)^{n_1+1}
    \prod_{k=1}^{{L-n_1-n_2}}
    \frac{\lambda_{j}^{(1)}-\lambda_{k}^{(2)}+i/2}
    {\lambda_{j}^{(1)}-\lambda_{k}^{(2)}-i/2},\quad j=1,\cdots,{L-n_1} , 
    \\
    \prod_{k=1}^{{L-n_1}}
    \frac{\lambda_{j}^{(2)}-\lambda_{k}^{(1)}+i/2}
    {\lambda_{j}^{(2)}-\lambda_{k}^{(1)}-i/2}&= 
    \prod_{k \ne j }^{{L-n_1-n_2}}
    \frac{\lambda_{j}^{(2)}-\lambda_{k}^{(2)}+i/2}
    {\lambda_{j}^{(2)}-\lambda_{k}^{(2)}-i/2}, 
    \quad j=1,\cdots,{L-n_1-n_2} \,.
  \end{aligned}
\end{equation}
The corresponding energies of the Hamiltonian with these boundary conditions
are given by Eq.~(\ref{ene1}), as before.

It is evident from this construction that the finite-size spectrum will be
different from that of the superspin chain only for sectors where the number
of fermions $n_1$ is even.  To see the effect on the scaling dimensions we
have to take into account that the Bethe state of highest weight in the
\emph{even} fermion sector of the $(p;q)$-multiplet is
parametrized by
\begin{equation}
  n_1=p\,,\quad n_2=2q
\end{equation}
rapidities in (\ref{bae1-bos}) for $p$ even rather than (\ref{pqnos}) which
still holds for $p$ odd.

The ground state of the spin chain with twist is the unique singlet $(0;0)$ as
for the superspin chain before.  The corresponding root configuration consists
of $L/2$ pairs of rapidities (\ref{bae1-pairs}).  
Unlike the situation in the superspin chain the ground state of the twisted
model has a strong finite-size dependence on $L$, see
Figure~\ref{fig:x_bos}(a):
\begin{figure}[h]
\includegraphics[width=0.65\textwidth]{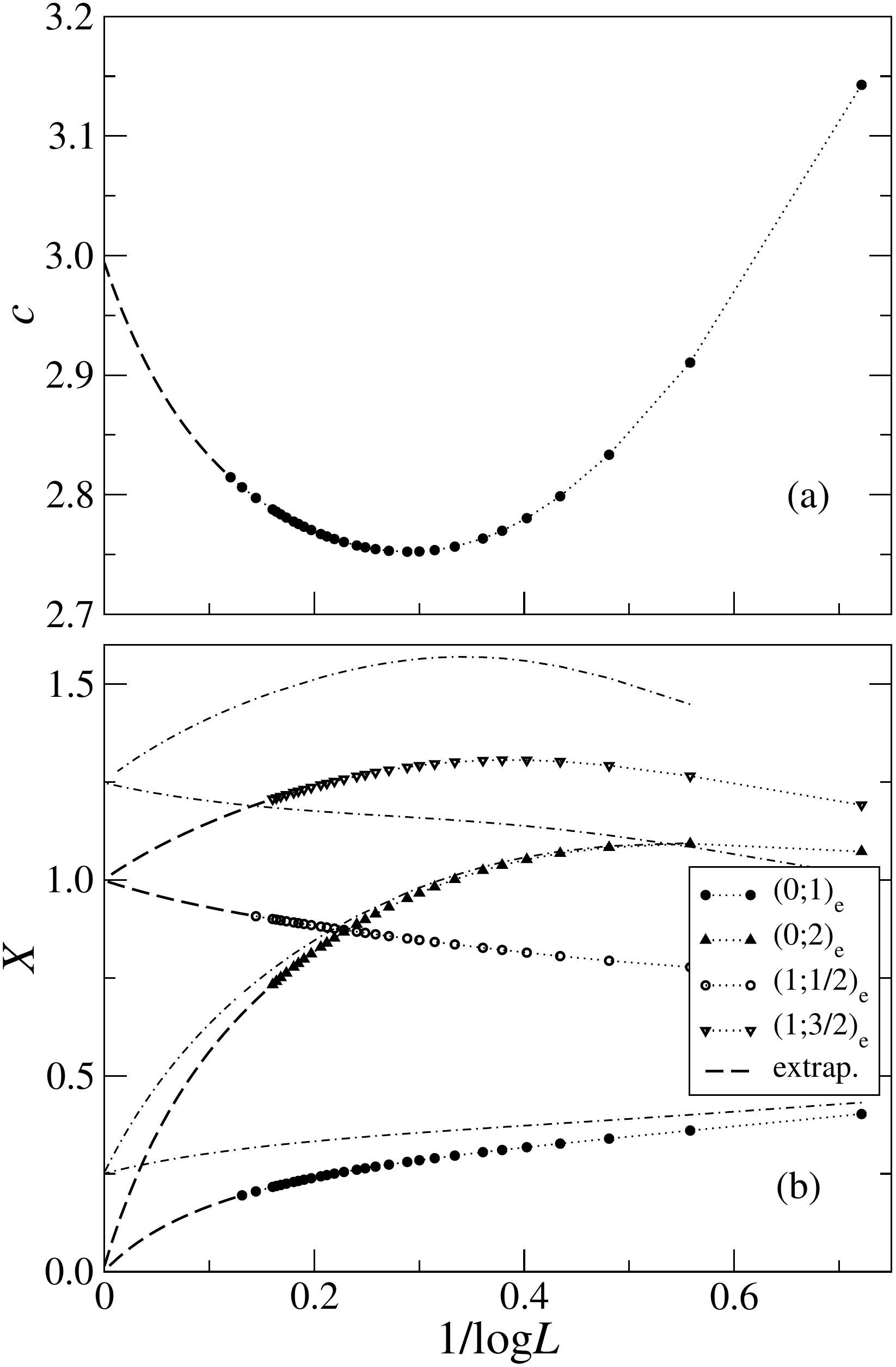}
\caption{(a) Central charge $c$ and (b) scaling dimensions $X$ extracted from
  the finite-size behaviour of the energies of the ground state and some low
  lying states $(p;q)_e$ with even fermion number of the spin chain subject to
  anti-periodic boundary conditions for the fermionic degrees of freedom.
  Dot-dashed lines in (b) indicate scaling dimensions extracted from the
  corresponding levels $(p;q)_o$ with an odd number of
  fermions. \label{fig:x_bos}}
\end{figure}
the additional phase in the Bethe equations (\ref{bae1-bos}) leads to a
central charge $c=3$ as noted before in Refs.~\onlinecite{MaNR98,JaRS03}.  In
addition there are subleading corrections to scaling which turn out to much
stronger than the $1/[\log{L}]^3$ behavior usually observed in isotropic spin
chains.  We stress that this is even in contrast to the finite-size behavior
found for other spin chains invariant by superalgebra such as ${OSp}(1|2)$ and
${OSp(2|2)}$.

In addition we have solved Eqs.~(\ref{bae1-bos}) for some low energy
excitations $(p;q)_e$ with even fermion number $n_1$.  Their root
configuration contains $n_2$ real roots $\lambda^{(1)}$ in addition to pairs
as in the ground state.
Our results show that the low energy spectrum of the spin chain with
anti-periodic boundary conditions for the fermionic states shows a similar
behaviour similar to that of the superspin chain, see
Figure~\ref{fig:x_bos}(b): the lowest excitations have fermion number $n_1=0$
with a continuous spectrum starting at scaling dimension $X_e=0$.  As
discussed above, the energies of states with \emph{odd} fermion $n_1$ number
are parametrized by the same rapidities as the corresponding $(p;q)$ multiplet
of the superspin chain.  Here the scaling dimensions, however, have to be
computed from (\ref{cft:fs}) relative to the new ground state which leads to a
shift $X_o=X_{(p;q)}+c/12$ as a consequence of the different effective central
charge.  As in the superspin chain the subleading corrections to the scaling
dimensions vanish as $1/\log L$.  The amplitudes of these terms display a
$q$-dependence which is clearly different from (\ref{fuseau-conj}), even for
the $X_o$ as a consequence of the logarithmic corrections to the central
charge.  Our data do not allow to quantify these amplitudes though: (much)
larger system sizes would be needed for an estimate which is beyond the
methods used in this work.


\begin{acknowledgments}
  MJM acknowledges the hospitality at the Institut f\"ur Theoretische Physik,
  Leibniz Universit\"at Hannover, where much of this work has been performed.
  Partial funding for this project has been provided by the Deutsche
  Forschungsgemeinschaft and the Brazilian Foundations FAPESP and CNPq.
\end{acknowledgments}

\newpage
\appendix
\section{Correspondence between the braid-monoid and the superalgebra
  realization of the intersecting loop model}
\label{app:braidmonoid}
The scattering rules of the Lorentz lattice gas and the configurations of the
intersecting loop are algebraically realized in terms of the generators of a
braid-monoid algebra first introduced by Brauer \cite{Brauer37}.  The braids
$\mathrm{B}_j$ correspond to the right mirrors, the monoids $\mathrm{E}_j$
represent the left mirrors while the identity $\mathrm{I}_j$ describes the
absence of a scatterer.  These operators acts on the sites $j=1,\cdots,L$ of
chain of length $L$ and satisfy a number of algebraic constraints:
\begin{itemize}
\item Braid relations
\begin{equation}
\mathrm{B}_j \mathrm{B}_{j \pm1} \mathrm{B_j} = 
\mathrm{B}_{j\pm1} \mathrm{B}_j \mathrm{B}_{j\pm1},~~ 
\mathrm{B}_j^2= \mathrm{I}_j,~~ \mathrm{B}_j \mathrm{B}_k =
\mathrm{B}_k\mathrm{B_j}~~ \mbox{for} \hspace{0.3cm} |j-k| \geq 2
\end{equation}

\item Monoid relations
\begin{equation}
\mathrm{E}_{j}\mathrm{E}_{j\pm1}\mathrm{E}_{j} = \mathrm{E}_{j},~~ 
\mathrm{E}_{j}^{2} = z\mathrm{E}_{j},~~ 
\mathrm{E}_{j}\mathrm{E}_{k} = \mathrm{E}_{k}\mathrm{E}_{j}~~ 
\mbox{for} \hspace{0.3cm}   |j-k| \geq 2 
\end{equation}
where ${z}$ is a free parameter representing the fugacity in the loop model
version.

\item Mixed relations
\begin{equation}
\mathrm{B}_{j}\mathrm{E}_{j} = \mathrm{E}_{j}\mathrm{B}_{j} = \mathrm{E}_{j},~~ 
\mathrm{E}_{j}\mathrm{B}_{j \pm 1} \mathrm{B}_{j} = \mathrm{B}_{j \pm 1}\mathrm{B}_{j}\mathrm{E}_{j \pm 1} 
= \mathrm{E}_{j}\mathrm{E}_{j \pm 1} 
\end{equation}
\end{itemize}
This algebra admits the construction of a one parameter family of integrable
models by means of the approach known as Baxterization \cite{Jones90}.  The
respective ${R}$-matrix is built out in terms of a weighted linear combination
of the generators,
\begin{equation}
  {R}_j=w_3\,\mathrm{I}_j+w_2\,\mathrm{E}_j+w_1\,\mathrm{B}_j\,.
\end{equation}

It turns out that the matrix ${R}_j$ fulfills the Yang-Baxter equation
provided that the weights $w_1$, $w_2$ and $w_3$ are sited on the projective
quadric,
\begin{equation}
  (w_1+w_2)w_3-\frac{2-z}{2}w_1w_2=0\,.
\end{equation}
This manifold can be parameterized with the help of one spectral parameter
$\lambda$ as follows,
\begin{equation}
  w_1= w_0\,,\quad
  w_2=\frac{w_0\lambda}{1-z/2-\lambda}\,,\quad
  w_3=w_0\lambda
\end{equation}
where $w_0$ is an overall normalization.  Note that this factor can be chosen
such that the weights configurations are interpreted as probabilities, i.e.\
$w_0=(z-2+2\lambda)/((z-2)(1+\lambda)+2\lambda^2)$.

It has been observed \cite{MaNR98} that for integer $z$ this algebra has a
realization in terms of a finite dimensional representation of the
superalgebra $OSp({m}|2{n})$ provided that
\begin{equation}
  z={m}-2{n}\,.
\end{equation}
In this formulation the braid operator becomes the graded permutation between
${m}$ bosonic and $2{n}$ fermionic degrees of freedom,
\begin{equation}
  \mathrm{B}_j  = \sum_{a,b=1}^{{m}+2{n}}(-1)^{p_ap_b} e_{ab}^{(j)} \otimes
  e_{ba}^{(j+1)}\,, 
\end{equation}
where $p_a$ is the Grassmann parity of the $a$-th degree of freedom
assuming values $p_a=0$ for bosons and $p_a=1$ for fermions. 

Similarly, the monoid operator can be written as,
\begin{equation}
  \mathrm{E}_{j} = \sum_{a,b,c,d=1}^{{m}+2{n}} \alpha_{ab} \alpha^{-1}_{cd}
  e_{ac}^{(j)} \otimes e_{bd}^{(j+1)}
\end{equation}
where the matrix elements $\alpha_{ab}$ are
\begin{equation}
  \alpha_{ab}=\begin{cases}{}
    (-1)^{1-{p_a}}\quad\mathrm{for~}a=b \cr
    +1\quad\mathrm{for~}a<b\mathrm{~with~}p_a=p_b=1\mathrm{~and~}
       \displaystyle \sum_{k=a+1}^{b}={n} \cr
    -1\quad\mathrm{for~}a>b\mathrm{~with~}p_a=p_b=1\mathrm{~and~}
       \displaystyle \sum_{k=b+1}^{a}={n} \cr
    0\quad\mathrm{otherwise} \cr
\end{cases}
\end{equation}
in terms of the gradings $p_a$.

We remark that for some specific grading ordering the matrix $\alpha$ can be
transformed into a block anti-diagonal structure.  For a generic example see
Ref.~\onlinecite{MaRa97a} and the main text, Eq.~(\ref{alph}), for the case of
$OSp(3|2)$.

\section{Finite dimensional representations of $OSp(3|2)$}
\label{app:osp}
As a consequence of the algebra inclusion $OSp(3|2) \supset SU(2)\oplus SU(2)$
one can use the labels of $SU(2)\oplus SU(2)$ irreps to classify the basis
states of an $OSp(3|2)$ irrep, see \cite{Jeugt84}:
except for the trivial representation $(0;0)$ the latter are characterized by
two integer or half integer numbers $(p\geq0;q\geq\half)$.  The quadratic and
quartic Casimir operators of $OSp(3|2)$ in terms of these numbers are
\begin{equation}
  \label{o32-casimir}
  I_2 = (p(p+1)+2q(1-2q))\,,\qquad
  I_4 = \frac14 I_2 \,(3p(p+1)+2(q+1)(2q-3))\,.
\end{equation}
For example, the five-dimensional fundamental representation $(0;\half)$ of
$OSp(3|2)$ carried by the local spins in the superspin chain decomposes as
\begin{equation}
  \label{o32-rep5}
  (0;\half) = (1\otimes0)\oplus(0\otimes\half)
\end{equation}
into $(\ell\otimes s)$-representations of $SU(2)\oplus SU(2)$.  States with
integer (half-integer) spin in the second factor are bosonic (fermionic).
Similarly, the first few of the other $OSp(3|2)$-representations with integer
$p$ (i.e.\ the only ones relevant for the $OSp(3|2)$ superspin chain) can be
decomposed as
\begin{equation}
\begin{aligned}
  (0;1) &= (0\otimes1)\oplus(1\otimes0)\oplus(1\otimes\half)\,,\\
  (0;q>1) &= (0\otimes q) \oplus (1\otimes q-1) 
                         \oplus (1\otimes q-\frac12) 
                         \oplus (0\otimes q-\frac32) \,,\\
  (p\geq1;\half) &=
       (p-1\otimes0)\oplus(p+1\otimes0)\oplus (p\otimes\half)\,,\\
  (1;1) &= (1\otimes1)\oplus(2\otimes0)\oplus(1\otimes\half)
           \oplus(2\otimes\half)\,,\\
  (p>1;1) &= (p\otimes0)\oplus(p\otimes\half)\oplus(p\otimes1)
           \oplus(p-1\otimes0)\oplus(p+1\otimes0)\\
           &\qquad\oplus(p-1\otimes\half)\oplus(p+1\otimes\half)\,,\\
\end{aligned}
\end{equation}
For $p\ge1$ and $q\ge\frac32$ with $p\ne2q-1$ the ``typical''
$OSp(3|2)$-representation $(p;q)$ has dimension $4(2p+1)(4q-1)$ and is an eightfold
pattern in the $SU(2)\oplus SU(2)$ decomposition
\begin{equation}
  \label{o32-typ}
\begin{aligned}
  (p;q) &= (p\otimes q) \oplus (p\otimes q-\frac12)
  \oplus (p-1\otimes q-\frac12)\oplus (p+1\otimes q-\frac12)\\
  &\qquad  \oplus (p\otimes q-1)\oplus (p-1\otimes q-1)\oplus (p+1\otimes q-1)
  \oplus(p\otimes q-\frac32)\,.
\end{aligned}
\end{equation}
For $p=2q-1$ some of the matrix elements of the $OSp(3|2)$ generators vanish which
allows to decompose (\ref{o32-typ}) into two ``atypical'' representations with
a fourfold $SU(2)\oplus SU(2)$ decomposition
\begin{equation}
  \label{o32-atyp}
    (p;q\equiv\frac12(p+1)) = (p\otimes q) \oplus (p\otimes q-\frac12)
  \oplus (p+1\otimes q-\frac12)\oplus (p+1\otimes q-1)\,.
\end{equation}
Note that the atypical representations together with the irreps $(0;\frac12)$,
$(1;1)$ and the trivial representation $(0;0)$ cannot be uniquely specified by
the eigenvalues of the Casimir invariants (\ref{o32-casimir}): for all of them
$I_2=I_4=0$.  Therefore, they can appear as parts of reducible but
indecomposable representations of $OSp(3|2)$ which are present in the spectrum of
the superspin chain as will be discussed below.

Based on these relations the Hilbert space of the isotropic $OSp(3|2)$
superspin chain for small $L$ can be decomposed into $OSp(3|2)$-multiplets
\begin{equation}
\label{o32-decomp}
\begin{aligned}
  (0;\half)^{\otimes 2}
     &= (0;0) \oplus (0;1) \oplus (1;\half)\,,\\
  (0;\half)^{\otimes3} &= (0;\half) \oplus (0;\frac32) \oplus (2;\half)
                     \oplus 2*\left[2*(0;\frac12)\oplus
                       (1;1)\right]_{\mathrm{ind}}
                     \,,\\
  (0;\half)^{\otimes4} &= (0;0) \oplus 6* (0;1) \oplus (0;2)
                   \oplus 6*(1;\frac12)  \oplus3*(1;\frac32)\\
                   &\quad
                   \oplus 3*(2;1) \oplus (3;\frac12)
                   \oplus 2*\left[2*(0;0)\oplus(1;1)\right]_{\mathrm{ind}}\,.
\end{aligned}
\end{equation}
By $\left[\ldots\right]_{\mathrm{ind}}$ we denote the combinations of atypical
representations in reducible but indecomposable ones.  The decompositions
(\ref{o32-decomp}) have been checked numerically.  Note that the trivial and
the fundamental representation of $OSp(3|2)$ appear exactly once in the tensor
product for even and odd $L$, respectively.\footnote{Other instances of these
  representations appear as parts of the indecomposables, though.}
Furthermore, all other multiplets appearing in (\ref{o32-decomp}) have even
dimension and contain the same number of bosonic and fermionic states. This
leads to the trivial partition function (\ref{partfun}) of the $OSp(3|2)$ vertex
model.

Reference states to be used in the algebraic Bethe ansatz of the superspin
chain with $L>1$ sites, each carrying the representation (\ref{o32-rep5}), can
be highest weight states of (either of) the $(8L-4)$-dimensional multiplets
$(L-1;\half)$ or $(0;\frac12 L)$.  

\newpage
%

\end{document}